# Efficient Cryptographic Substitution Box Design Using Travelling Salesman Problem and Chaos

Musheer Ahmad[a*], Nikhil Mittal[a], Prerit Garg[a], Manaff Mahtab Khan[a]

[a]*Department of Computer Engineering, Faculty of Engineering and Technology,*
*Jamia Millia Islamia, New Delhi-110025, India*

**Abstract**

Symmetric encryption has been a standout amongst the most reliable option by which security is accomplished. In modern block symmetric ciphers, the substitution-boxes have been playing a critical role of nonlinear components that drives the actual security of ciphers. In this paper, the travelling salesman problem and piece-wise linear chaotic map are explored to synthesize an efficient configuration of 8×8 substitution-box. The proposed anticipated design has the consistency which is justified by the standard performance indexes. The statistical results manifest that the prospective substitution-box is cryptographically more impressive as compared to some recent investigations.

## 1. Introduction

With an increase of information traffic on internet, security of information has become one of the prime concerns for the military, academicians, researchers, online business, technocrats, etc. Encryption is one of the reliable technique by which security is assured. The modern cryptographic techniques are majorly classified into two fields; viz. symmetric-key encryption and asymmetric-key encryption. The aspect at which these fields differ is the way the keys are exchanged during the communication. Modern block ciphers like AES, DES, IDEA, Twofish, etc., are symmetric key encryptions. A strong cryptosystem must satisfy the properties of confusion and diffusion. The diffusion ensures that the plain-text cannot be discerned from the cipher-text by hiding the relationship between them. Whereas, confusion hides the intermediary within key and encrypted data ensuring that the key cannot be guessed by any unauthorized intruder. In block ciphers, the property of confusion is achieved by employing substitution-boxes (S-boxes) and diffusion is imparted through the usage of permutation-boxes (P-boxes) which are linear in nature [1]. Let $g: B^n \rightarrow B^m$ denotes an $n \times m$ S-box which maps $n$-bits input values to $m$-bits outputs. In cases where $n = m$, the S-box are bijective in nature if all outputs occur exactly once. The S-boxes are the only prominent nonlinear components that decide the security strength of block ciphers. They are designated to alter the plain-text to lend high nonlinearity and strength against attacks. An S-box with low nonlinearity scores considered as cryptographically weak and can compromise the security of cryptosystem. Therefore, it has been a challenge in research to construct cryptographically potent substitution-boxes that can impart high non-linearity and security to cryptosystem.

* Corresponding author. E-mail: musheer.cse@gmail.com, Tel: +91-112-698-0281
E-mail: nkmittal4994@gmail.com (N. Mittal), prerit2010@gmail.com(P. Garg), manaffkhan13@gmail.com (M.M. Khan)

The chaos-based cryptography has attracted the attentions of researchers, academicians and scientists in areas of mathematics, computer science, engineering, etc. Chaos has been considered as fitting to design security primitives on account of its enticing lineaments such as being sensitive to initial conditions, pseudo-random-like, long periodic and high ergodicity [1, 2]. It has been exhaustively explored to construct methods for text, image, video, audio encryptions, hash functions, S-boxes, authentications, data hiding, etc. This paper proposes to present novel approach to synthesis efficient S-box using travelling salesman problem and PWLCM chaotic map. The performance measures are quantified to evaluate strength of proposed S-box. The results of statistical analyses are compared with some recently synthesized S-boxes and found to possess better virtues.

## 2. Proposed Design

### 2.1. Traveling Salesman Problem

The travelling salesman problem was formulated in 1830s by mathematicians Hamilton and Kirkman [3]. The TSP is a standout amongst the most considered combinatorial advancement issues from NP set. The TSP problem is to discover efficient hamiltonian tour in weighted graph of cities a salesman can take. The solution for TSP appreciates wide pertinence in an assortment of practical areas of study. Some of the finest applications of TSP include scheduling problems, DNA sequencing, bus routing problem, etc.

## 2.2. Piece-Wise Linear Chaotic Map

Among dynamical systems, the piece-wise linear chaotic map is 1D system is widely contemplated chaotic systems whose state equation governs as [2]:

$$x(n+1) = pwlcm(x, p) = \begin{cases} \dfrac{x(n)}{p} & 0 < x(n) \le p \\ \dfrac{1 - x(n)}{1 - p} & p < x(n) < 1 \end{cases} \quad (1)$$

Where $x \in (0, 1)$ for all $n \ge 0$ is variable, $p$ represents control parameter, $n$ is count of iterations and $x(0)$ defines the map's initial condition. The PWLCM shows chaotic demeanour for all $p \in (0,1)$.

## 2.3. Algorithm

The proposed algorithm is designed to employ TSP and chaos in novel manner. Like other S-box proposals, the anticipated algorithm also has the features of simplicity and low computational overheads. The steps of algorithm are follows as:

**Step 1.** Generate an initial S-box as:

    1.1 Let $S = \{\}$ be an empty set, take initial values of $x$ and $p$.

    1.2 Execute PWLCM map for 1000 times and throw away the values excluding the last to remove the transient effect of map.

    1.3 Repeat steps 1.4 to 1.6 while length of $S \ne 256$

    1.4 Further execute map and save current $x$ variable

    1.5 Extract $t$ as $t = floor(x \times 10^{10}) mod(256)$

    1.6 IF $t \in S$ then goto step 1.4, ELSE add $t$ to $S$ and goto step 1.4

    $S$ holds initial 8×8 S-box and contains 256 values which are considered as nodes of graph.

**Step 2.** Decompose $S$ linearly into 32 complete sub-graphs, each containing 8 nodes and 8×(8-1)/2 = 28 edges.

**Step 3.** Further iterate PWLCM map to assign weights $\in [1, 255]$ to edges of sub-graphs.

**Step 4.** Apply TSP on each sub-graph and record the hamiltonian tours.

**Step 5.** Update S-box $S$ according to the efficient tours obtained.

**Step 6.** Process $S$ using 33<sup>rd</sup> complete sub-graph as:

    6.1. Select eight middle elements of $S$ as nodes to make next sub-graph.

    6.2. Further iterate PWLCM map to assign weights to edges

    6.3. Apply TSP and update $S$ according to the tour.

**Step 7.** Lastly, process *S* using 34[th] complete sub-graph as:

    7.1. Select eight last elements of *S* as nodes to make next sub-graph.

    7.2. Further iterate PWLCM map to assign weights to edges

    7.3. Apply TSP and update *S* according to the tour.

The final configuration of S-box obtained with proposed approach is listed below in Table 1.

Table 1: Proposed substitution-box

| 236 | 27  | 24  | 98  | 149 | 101 | 165 | 201 | 226 | 80  | 9   | 89  | 31  | 181 | 188 | 4   |
|-----|-----|-----|-----|-----|-----|-----|-----|-----|-----|-----|-----|-----|-----|-----|-----|
| 95  | 191 | 100 | 138 | 99  | 125 | 13  | 139 | 87  | 79  | 60  | 64  | 131 | 67  | 97  | 32  |
| 14  | 186 | 93  | 58  | 122 | 155 | 135 | 37  | 103 | 180 | 86  | 48  | 179 | 110 | 0   | 30  |
| 147 | 145 | 166 | 156 | 70  | 21  | 248 | 163 | 207 | 178 | 159 | 224 | 57  | 112 | 51  | 210 |
| 120 | 94  | 243 | 39  | 190 | 69  | 241 | 211 | 20  | 253 | 208 | 136 | 153 | 56  | 198 | 197 |
| 77  | 189 | 29  | 218 | 68  | 42  | 17  | 246 | 157 | 109 | 175 | 192 | 183 | 148 | 53  | 113 |
| 118 | 245 | 116 | 43  | 130 | 22  | 12  | 212 | 134 | 213 | 162 | 229 | 151 | 1   | 34  | 36  |
| 111 | 52  | 124 | 75  | 203 | 254 | 126 | 81  | 249 | 146 | 62  | 73  | 247 | 40  | 240 | 132 |
| 167 | 114 | 72  | 142 | 25  | 44  | 214 | 242 | 49  | 123 | 237 | 225 | 23  | 137 | 85  | 152 |
| 7   | 121 | 255 | 46  | 19  | 160 | 220 | 195 | 10  | 133 | 177 | 47  | 127 | 196 | 169 | 63  |
| 141 | 91  | 194 | 185 | 204 | 92  | 216 | 215 | 234 | 107 | 223 | 82  | 108 | 219 | 59  | 6   |
| 140 | 16  | 193 | 88  | 71  | 161 | 154 | 50  | 28  | 106 | 61  | 129 | 227 | 54  | 76  | 170 |
| 168 | 206 | 45  | 35  | 18  | 228 | 184 | 26  | 200 | 83  | 174 | 105 | 238 | 171 | 90  | 217 |
| 8   | 38  | 187 | 104 | 176 | 41  | 65  | 239 | 84  | 11  | 128 | 172 | 158 | 74  | 235 | 250 |
| 231 | 66  | 2   | 232 | 199 | 230 | 143 | 222 | 102 | 164 | 205 | 78  | 117 | 202 | 144 | 3   |
| 251 | 119 | 173 | 96  | 209 | 33  | 115 | 15  | 150 | 233 | 182 | 244 | 5   | 55  | 252 | 221 |

## 3. Statistical Results

In this section, the strength of substitution boxes is quantified against notable performance indexes such as bijectiveness, degree of nonlinearities, avalanche effect and differential uniformity. These statistical parameters are evaluated to annotate its pertinency for block ciphering. Further, the outcomes are adjudged against some recently investigated S-boxes.

### 3.1. Bijectiveness

An 8×8 S-box is said to be bijective, if it is a one-to-one and onto mapping from input vector to the output vector or if S-box has all unique elements in the range [0, 255]. Since, the S-box depicted in Table 1 has unique entries in the specified range. It can be claimed that the proposed S-box satisfies the bijectiveness.

### 3.2. Nonlinearity

For a cryptographic *n*-bit Boolean function *f*, its nonlinearity symbolizes the degree of dissimilarity between *f* and *n*-bit affine function most similar to *f*. It is the least hamming distance between the vector representing function's truth table and the set of all *n*-bit affine functions. A function having high minimum hamming distance is said to have high nonlinearity. High nonlinearity provides resistance to linear approximation attacks [6, 7]. Contrarily, Boolean function with low nonlinearity can be approximated by some affine function. The nonlinearity of a Boolean functions f is accounted as:

$$N_f = 2^{n-1}(1 - 2^{-n} \max |S_{\langle f \rangle}(w)|) \quad \text{and} \quad S_{\langle f \rangle}(w) = \sum_{w \in GF(2^n)} (-1)^{f(x) \oplus x.w} \quad (2)$$

Where, $S_{(f)}(w)$ is the Walsh spectrum of function $f$ and $x.w$ denotes the dot-product of $x$ and $w$. The nonlinearity scores for the eight Boolean functions of proposed S-box are 108, 110, 110, 108, 106, 106, 106, 106 providing an excellent statistics like 106 as minimum, 110 as maximum and 107.5 as average score which are undoubtedly more influential than scores of other S-boxes investigated in [4-7]. Hence, the proposed S-box offers better nonlinearity, security and resistance to linear attacks.

Table 2: Comparison of nonlinearities of some recent 8×8 substitution-boxes

| S-box | $n_1$ | $n_2$ | $n_3$ | $n_4$ | $n_5$ | $n_6$ | $n_7$ | $n_8$ | min | max | mean |
|---|---|---|---|---|---|---|---|---|---|---|---|
| Proposed | 108 | 110 | 110 | 108 | 106 | 106 | 106 | 106 | 106 | 110 | 107.5 |
| Ahmad *et al.* [1] | 108 | 106 | 106 | 106 | 106 | 110 | 106 | 108 | 106 | 110 | 107 |
| Özkaynak *et al.* [4] | 104 | 100 | 106 | 102 | 104 | 102 | 104 | 104 | 100 | 104 | 103.3 |
| Khan *et al.* [5] | 108 | 102 | 100 | 104 | 104 | 102 | 98 | 106 | 98 | 108 | 103 |
| Gondal *et al.* [6] | 98 | 100 | 106 | 104 | 106 | 100 | 106 | 104 | 98 | 106 | 103 |
| Belazi *et al.* [7] | 102 | 106 | 104 | 106 | 108 | 106 | 106 | 104 | 102 | 108 | 105.25 |

*3.3. Strict Avalanche Criteria*

The concept of strict avalanche criteria (SAC) was initially presented by Webster and Tavares [8]. Accordingly, a change in one of the input bits must lead to change in half of output bits. Thus, when SAC is satisfied, a small change in the input leads to a significant change in the output. This is desirable because the resulting output vector then appears to be highly random, and no pattern can be recognised by slight modification in the input vector. The quantified SAC for proposed S-box is 0.5036 which is acceptable statistics since it is quite close to ½.

*3.4. Differential Uniformity*

An S-box ought to have as low differential uniformity (DU) as could be allowed. An input differential $\Delta x$ ought to map unambiguously to an output differential $\Delta y$, guaranteeing a uniform mapping likelihood. The maximum value of DU should be as low as possible to thwart the differential attacks [1, 4]. The DU of an S-box is a measure of differential consistency and calculated as:

$$DU = \max_{\Delta x \neq 0, \Delta y} \left( \#\{x \in X \mid S(x) \oplus S(x \oplus \Delta x) = \Delta y\} \right) \tag{3}$$

Where, *X* is the set of all possible input values of S-box. The proposed S-box offers a max DU of 10 which is superior than max DU score of 12 of S-boxes investigated in [5,6,7].

## 4. Conclusion

In this paper, an efficient 8×8 substitution-box synthesis scheme is presented which is based on the concept of travelling salesman problem and chaotic map. A novel idea is explored to yield cryptographically proficient setup of substitution-box. The performance excellency, consistency and acceptability of the proposed scheme and S-box is defended by the standard statistical outcomes. The experimental results show that the anticipated substitution-box is cryptographically more impressive when contrasted with some recently investigated S-boxes.